\documentclass[12pt]{iopart}
\usepackage{epsfig}
\begin{document}
\title{Entanglement dynamics of a two-particle scattering in pulsed sinusoidal potentials} 

\author{F Buscemi$^{1,2}$, P Bordone$^{3,2}$ and A Bertoni$^2$}

\address{$^1$ ARCES, Alma Mater Studiorum, University of Bologna, Via Toffano 2/2, 40125 Bologna, Italy}
\address{$^2$ S3 Research Center, CNR-INFM, Via Campi 213/A, I-Modena 41100, Italy }
\address{$^3$ Dipartimento di Fisica, Universit\`{a} di Modena
e Reggio Emilia,I-41100 Modena, Italy}

\eads{\mailto{fabrizio.buscemi@unimore.it} }

\begin{abstract}
We study by means of time-dependent numerical simulations the behavior of the entanglement  stemming from the Coulomb scattering  between two charged particles  subject to a pulse of sinusoidal potential.  We show that the splitting of the spatial wavefunction brought about by the interaction  with the potential pulse plays a key role in the appearance of quantum correlation, thus leading  under specific conditions to a peculiar behavior. The dependence of the final  entanglement  upon the physical parameters describing the pulse  is discussed. Our results can be applied to a number of physical systems, such  as electron-electron scattering in semiconductors or cold-ions dynamics in external fields.
\end{abstract}

\submitto{\JPA} 

\section{Introduction} 

Quantum entanglement between two spatially separated particles
can be created as a consequence of their non local mutual interactions \cite{Hars,Law}. In particular, entangled states are produced from the collision between two charged particles
interacting via Coulomb potential.  Indeed, after a
scattering the two-particle system is in general described by a
two-particle state that is not separable in two single-particle pure
states.  This entanglement building up is an intrinsically dynamical
process and its analysis is not only useful to understand the nature
of the scattering process itself, but it can also contribute
to design quantum information processing devices.  In
fact, on one hand, controlled entanglement has been recognized as the
fundamental resource for quantum computation and communication \cite{Peres}, on the
other hand entanglement with the environment (i.e. decoherence)
represents the main threat to the proper functioning of a feasible
quantum computer \cite{Giuli}.
For those reasons, the study of the entanglement dynamics in
scattering events  has became more and more
relevant in recent years\cite{Tal,bus2,Gun,Costa}
and different proposals to produce entangled states between charged
carriers in solid-state systems have been presented\cite{berbor,Oli,Ram}.

In this work, we analyze the time evolution of the entanglement during the
scattering of two particles  in presence of an external  periodic potential.
Our model and results are representative of a
broad variety of situations in which  two interacting carriers
are constrained in a quasi-1D domain and a sinusoidal-like potential is
present. Such kind of systems are of interest in different areas of physics,
such as condensed matter\cite{Devereaux}, quantum optics\cite{Scott}
and astrophysics\cite{Fro}.  In particular a great attention has been
devoted to the scattering of an electron beam by a standing wave of
light (the so-called Kaptiza-Dirac effect)\cite{Kap}, stemming from
the possibility of using such a system to investigate the wave nature
of electrons.  The latter model of matter-field interaction also
raises conceptual and theoretical issues about the momentum exchange
between electrons and electromagnetic radiation\cite{Bate,Min, Efre,
Efre2}, which can be generalized to fields other than quantum optics.

% As a prototype system, we consider a semiconductor quantum wire,
% eventually in the presence of SAWs\cite{Barnes,rosini,Shilton,Taly,
% Cunn, Ebbeke}.  The SAWs are lattice vibrations that propagate
% through a semiconductor structure as longitudinal waves and
%  are produced in the experimental apparatus by high frequency AC
% transducers.  In GaAs/AlGaAs devices, they couple to charge carriers
% through a piezoelectric mechanism and
% can be modelled as a sinusoidal
% travelling electrical potential which traps the carriers in its moving
% minima.  A number of deviced exploiting SAW-electron interaction have
% been designed, realized and also proposed as basic building blocks for
% quantum computing applications, where the use of SAW has been
% shown to constitute a highly controllable mean to inject and drive
% electrons along quantum wires\cite{rosini}.  In fact, although the SAW technology
% was originally introduced in the context of metrological application
% for defining a new standard of electric current, it has been recognized as
% an important resource to improve the functionality of semiconductor
% quantum logic gates\cite{rosini}.

The focus of the present work is on the creation of quantum
entanglement between two interacting electrons, in the presence of a
further external periodic potential.  The two particles are explicitely considered as
indistinguishable and have the same spin. Specifically, we simulate
numerically a two-particle scattering in a 1D structure
and determine how the tailoring of a standing pulse of sinusoidal
potential is able to affect electron-electron correlation.  
  Here we intend to move a
step forward in the investigation of the entanglement formation in
electron-electron scattering processes, recently addressed by many
works\cite{Tal, bus2,Gun, bus}.  We analyze how the momentum exchange
between the particles affects the entanglement arising in the binary
collision, and how the inclusion of a  potential oscillating
in space modifies the entanglement dynamics.  To
this aim we tackle the problem by solving numerically the
time-dependent two-particle Schr\"odinger equation and by computing,
at each time step, the bipartite entanglement.  Such an approach
allows us to investigate also the case of a non-adiabatic switching
on/off of the potential, that would make analytical techniques not
applicable.

% We also address another important aspect related to electron
% transport assisted by SAW.  In fact, such a system is commonly
% investigated in the single-particle approximation, thus neglecting the
% quantum correlations.  Here we want to verify whether this can be
% really considered a good approximation.  Moreover we analyze the role
% played by the sinusoidal potential induced by the SAW in preventing
% the spreading of the wavefunction and in reducing undesired reflection
% effects.

The paper is organized as follows.  In Sec.~\ref{intro}, to better
introduce our model system, we study the dynamics of a single free
electron that is subject to a sinusoidal potential only for a small
time interval. % We will refer to this  as ``pulsed potential''
 In
Sec.~\ref{intro2} we evaluate  the entanglement generated in a
collision between two electrons  subject to a pulsed potential. %  then
% between two electrons localized in a steady sinusoidal potential,
% finally between two electron in the latter condition, when the
% external potential is removed for a short time interval.
We comment
on the results and draw final remarks in Sec.~\ref{intro5}.

%%%%%%%%%%%%%%%%%%%%%%%%%%%%%%%%%%%%%%%%%%%%%%%%%%%%%%%%%%%%%%%%%%%%%%%%%%

\section{Single-particle system} \label{intro}
In this section we study the dynamics of a free electron propagating
in a quasi 1D system and subject to a single pulsed sinusoidal
potential.  Our aim is to investigate the role played by the sine-like
time-dependent potential in the time evolution of a simple
single-particle wave function.  The results will be of help in
understanding the two-particle dynamics of the following section.

The single particle is described at the initial time $t_0=0$ by a
minimum uncertainty wave-packet, with the following wave function
\begin{equation} \label{rosi}
\psi(x,t_0)= \frac{1}{(\sqrt{2 \pi}\sigma)^{1/2}}
\exp{\left( -\frac{(x-x_{0})^{2}}{4\sigma^{2}}+
\frac{i}{\hbar}p_{in}\cdot x\right)}
\end{equation}
where $\sigma$ is the mean dispersion in position and $p_{in}$ the
initial momentum.

The particle  feels  a pulsed sine-like potential, and the
Hamiltonian of the system takes the form
\begin{equation}\label{rest}
H_{on-off}(x)= - \frac{\hbar^{2}}{2m}\frac{\partial^{2}}{\partial x^{2}}+A \sin(k_0 x)
\Theta(t-t_{on})\Theta(t_{off}-t)
\end{equation}
where $m$ is the particle  mass, $A$ the amplitude of the potential  oscillation, $k_0$ its
the wavenumber and $\Theta$  the Heaviside function, with $t_{on}$ and $t_{off}$
the times of turning on and off of the potential, respectively.

We note that the time-dependent Hamiltonian of Eq.~(\ref{rest}) has
the same form of the one used to describe the scattering of an
electron by a standing light wave.  Such an interaction has been
widely studied both from the theoretical and the experimental points
of view\cite{Bate,Min, Efre, Efre2}, beginning with the the original
work by Kaptiza and Dirac\cite{Kap}.  In the literature, the standing
light electromagnetic potential is taken as a superposition of two
counterpropagating travelling waves of identical frequency and the
characteristic times of the scattering process are assumed to be much
longer than the period of the wave. As a consequence, the
approximation of time-average Hamiltonian can be used, thus leading to
the so-called ponderomotive potential oscillating in
space\cite{Bate,Efre2}.

Some peculiar aspects of the interaction between the electron and the
sinusoidal wave can be understood by analyzing the dynamics of the
single-particle wavefunction in the momentum representation
$\phi(k,t)$, the latter being the Fourier transform of the real-space
wave function $\psi(x,t)$, with $k=p/\hbar$.  At the initial time,
$\phi(k,t_0)$ is given by a Gaussian wavepacket with a mean dispersion
$1/\sigma$ centered around $k_{in}=p_{in}/\hbar$.  Let us now introduce
the wavefunction $\varphi(k,t)$ in the interaction picture
\begin{equation}\label{jers1}
\phi(k,t)= \exp{\left(-i\,\frac{\hbar k^2}{2m} t \right)}\varphi(k,t).
\end{equation}
Its time evolution can be evaluated by inserting Eq.~(\ref{jers1}) into
the Schr\"{o}dinger equation of the system, whose Hamiltonian is given
in Eq.~(\ref{rest}) for $t_{on} \le t \le t_{off}$.  Straightforward
calculations lead to the following recurrence relation\cite{Efre,Efre2}:
\begin{eqnarray}\label{jers2}
\frac{\partial}{\partial t}\varphi(k,t)=& -&\frac{A}{2 \hbar}
\exp{\left[i\left(\frac{\hbar k k_0}{m}-\frac{\hbar  k_0^2}{2m}\right)t
\right]}\varphi(k-k_0,t)          \nonumber        \\
&+&\frac{A}{2\hbar}\exp{\left[-i\left(\frac{\hbar k k_0}{m}+
\frac{\hbar  k_0^2}{2m}\right)t\right]}\varphi(k+k_0,t).
\end{eqnarray}
This represents the dynamical equation for $\varphi(k,t)$ due to the
interaction between the particle and the sinusoidal potential.  The two
terms appearing in the rhs of the Eq.~({\ref{jers2}}) have a clear
physical meaning: the interaction with a potential oscillating in space
allows the particle to change its momentum by an integer number of
$\pm \hbar k_0$.  This consideration will turn out to be fundamental to
explain our results.

\begin{figure}[h]
  \begin{centering} \includegraphics*[width=\linewidth]{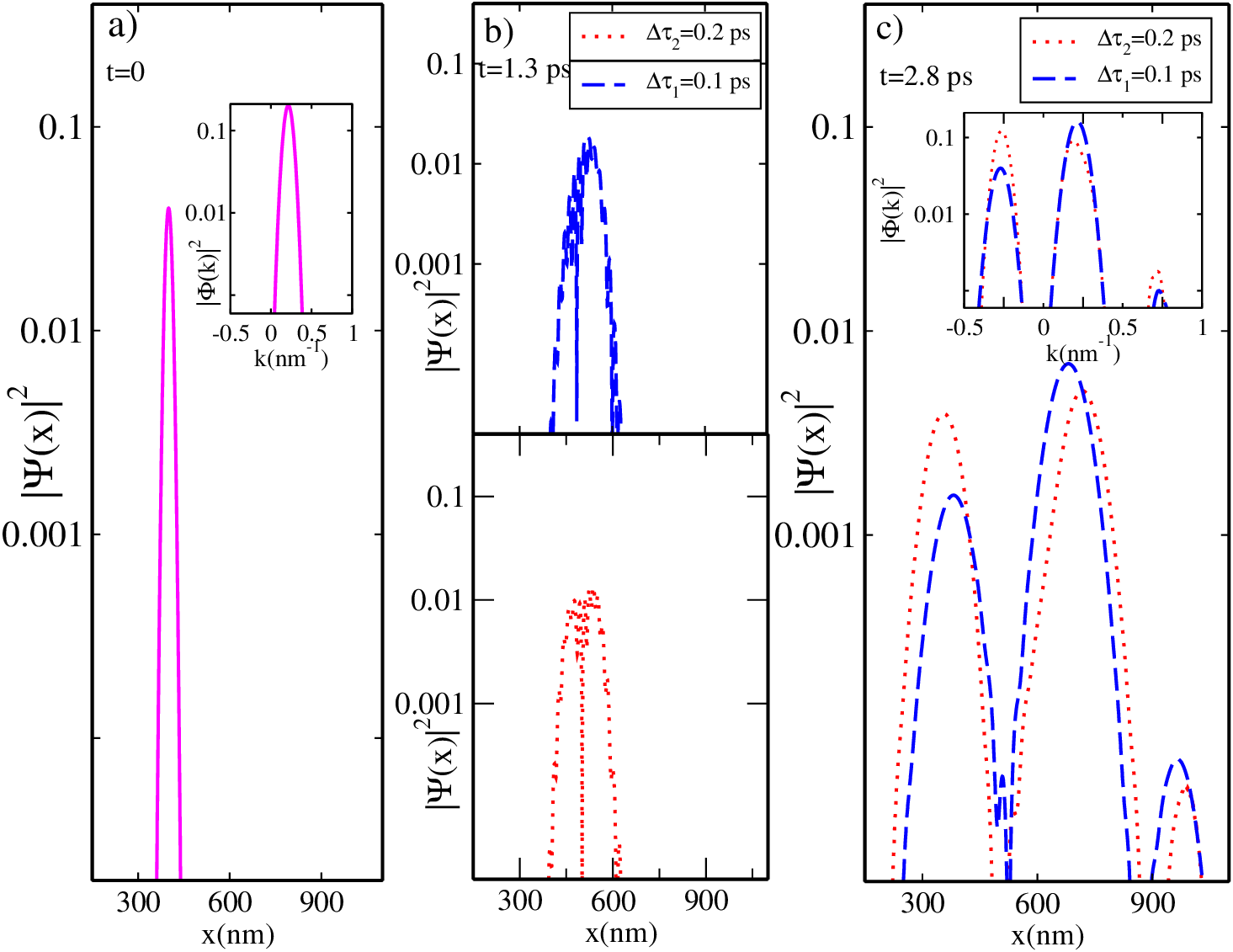}
  \caption{\label{fig:con1} (a) Square modulus of the  single-electron 
  wave function $\psi(x)$ at the initial time.
 The inset shows the square modulus of its
  Fourier transform $\phi(k,0)$.  (b) Top panel:
  square modulus of the electron wavefunction $\psi(x)$ (dashed line)
  at $t=$1.3 ps, for a pulse duration  $\Delta\tau_1= 0.1$ ps. % (The external potential is sketched by the thin solid line).
  Bottom
  panel: same as top for $\Delta\tau_2$= 0.2 ps. In both cases
  $t_{on} =1 $ ps. % Note that, unlike (a) and (c) panels, here the scale of the ordinata is taken linear in order to plot the sinusoidal potential 
(c) Comparison of the square modulus of the two electrons wavefunctions at $t=$2.8 ps, i.e. after the pulse, for $\Delta\tau_1$ (dashed line) and
  $\Delta\tau_2$ (dotted line). The inset displays their Fourier transforms. In these numerical calculations we have taken  $m=2.91\times$  $10^{-31}$ Kg,
$\sigma$=10 nm, $k_0=0.46$ nm, $k_{in}=0.205$ nm and $A$=6.11 meV. They are the  physical parameters which could describe electron  transport in low dimensional semiconductor structures. Note the logarithmic scale, adopted for clarity.}  \end{centering}
\end{figure}

Under some specific conditions, approximate analytical solutions of
Eq.~(\ref{jers2}) can be obtained\cite{Min, Efre, Efre2}.  However,
for the sake of generality, we face the problem numerically by
solving the time-dependent Schr\"{o}dinger equation for the electron
real-space wavefunction by means of a Crank-Nicholson finite
difference scheme.  In Fig.~\ref{fig:con1} we report the square
modulus of the wavefunction $\psi(x,t)$ given in Eq.~(\ref{rosi}) at
three different time steps and for two different pulse lengths
$\Delta\tau=t_{off}-t_{on}$, namely $\Delta\tau_1$=0.1 ps and  $\Delta\tau_2$=0.2 ps.  In both cases
the sine potential is turned on at $t_{on}= 1$~ps. The sinusoidal potential is included in the time evolution of the system for the duration $\Delta\tau$ of the pulse.
From panel (b) we observe that shortly after $t_{off}$ the wavepacket
is not described by a smooth function anymore, but exhibits rapid
oscillations. Their appearance can be ascribed
to the instantaneous potential switching off
giving rise to a 
 large energy uncertainty. However  such oscillations are less pronounced  at longer times  owing to the natural spreading of the wavefunction. The latter results to be split in three peaks, as can be seen from
panel (c).  In order to understand this behaviour we need to analyze
the dynamics of the momentum wave function $\phi(k,t)$, whose square
modulus, after the pulse of sinusoidal potential, shows three peaks
(see the inset of panel (c) of Fig.~\ref{fig:con1}): one is still
centered in $ k_{in}$, while the other two are centered in
$k_{in}+k_{0}$ and $k_{in}-k_{0}$. This suggests us that, for 
 the sinusoidal pulses  of length  $\Delta\tau_1$ and  $\Delta\tau_2$, the
particle is scattered by the potential and its momentum can be
possibly increased or decreased by $\hbar k_{0}$, in perfect agreement
with the prediction of Eq.~(\ref{jers2}).  % In this spirit we can
% assume that, for the case of a series of periodic pulses of the
% sine-like interaction, it should be expected a quantum chaotic
% behaviour, characterized by the so-called \emph{dynamical localization}
% effect, that is the non diffusive but exponentially localized momentum
% distribution~\cite{Frasca, Gard}.  It is worth noting that the
% momentum gain or loss in our model depends strictly upon the
% duration of the pulse $\Delta\tau$.
Specifically, the $\Delta\tau$ used in this work can be considered
sufficient to induce a variation of $\hbar k_{0}$ in the electron
momentum, while the transfer of larger multiples of $\hbar k_{0}$ has
a very small probability, as revealed by the negligible amplitude of
the corresponding peaks in the momentum representation (not included in Fig.~\ref{fig:con1}).

Thus, for pulses of duration $\Delta\tau_1$ and  $\Delta\tau_2$, the splitting of
the wavefunction into three peaks after $t_{off}$, has an immediate
physical interpretation. The central peak gives the free evolution of
the electron with momentum $p_{in}$ while the other two describe the
single-particle free dynamics with momentum $p_{in}+\hbar k_{0}$, the
``accelerated component'' and $p_{in}-\hbar k_{0}$, the ``reflected
component''. For both pulse lengths the peak corresponding to the acceleration is smaller than the one giving the   reflection. This asymmetry
can be related to the form of the dynamical equation for  the momentum wave
function $\phi(k,t)$
given in the  Eq.~(\ref{jers2}). Here, the two terms
describing the addition and the subtraction of the momentum $k_0$
 contain  the factors $\exp{\left[-i\left(\frac{\hbar k k_0}{m}+
\frac{\hbar  k_0^2}{2m}\right)t\right]}$ and $\exp{\left[i\left(\frac{\hbar k k_0}{m}-\frac{\hbar  k_0^2}{2m}\right)t
\right]}$: the former oscillates in time  more
rapidly than  the latter and therefore its integration gives  a smaller 
contribution to $\phi(k,t)$.

%Ho tolto questo pezzo: puo' essere oggetto di facile critiche
% However, we find remarkable differences in the splitting
% of the single-particle wavepacket when $\Delta \tau$ changes from 0.1
% ps to 0.2 ps.  In fact, for $\Delta\tau = 0.1$ the central peak is
% higher than the one of the ``reflected component'', while for
% $\Delta\tau = 0.2$ they are very similar.

%%%%%%%%%%%%%%%%%%%%%%%%%%%%%%%%%%%%%%%%%%%%%%%%%%%%%%%%%%%%%%%%%%%%%%%%%%%%%

\section{Electron-electron entanglement}\label{intro2}
We now focus on the entanglement created in a two-electron scattering.
The two particles interact via the Coulomb repulsion and are subject
to a sinusoidal pulse, as described in the previous section.  
Since the entanglement formation in 1D- and 2D-scattering events
between two unbound and/or trapped particles has been recently
investigated\cite{Tal,bus2,bus}, it appears of interest to study the
building up of quantum correlations in such systems when a
time-dependent external potential is introduced.

Here, the two interacting particles run in opposite directions along a
1D structure. The external sinusoidal potential is switched on at
$t_{on}$ and for a time interval $\Delta\tau$. The Hamiltonian of the
system reads
\begin{eqnarray}
  H(x_{a},x_{b})=  H_{on-off}(x_a)+H_{on-off}(x_b)+
\frac{e^{2}}{\epsilon \sqrt{\left(x_{a}-x_{b}\right)^{2}+d^{2}}} 
\end{eqnarray}
where $ H_{on-off}$ is the single-particle Hamiltonian given in
Eq.~(\ref{rest}), $\epsilon$ is the  dielectric constant and
$d$ represents the cut-off  the Coulomb interaction.\footnote{The values of the physical parameters used in the calculation refer to a 1D scattering model in a silicon quantum wire of width $d$,but the results obtained are representative
of a more general behaviour.}

The two carriers have the same spin (up) and are obviously
indistinguishable,  so that the quantum state
describing the system is given by
\begin{equation} \label{mefi}
|\Psi\rangle=\frac{1}{\sqrt{2}}\bigg(|\psi \,\phi\rangle
- |\phi \,\psi\rangle \bigg)|\!\!\uparrow \uparrow \rangle.
\end{equation}
where both the wavefunctions corresponding to the states
$|\psi\rangle$ and $|\phi\rangle$ are of the type defined in
Eq.~(\ref{rosi}).  The initial spread of the wavepackets and the
distance between their centers are such that the Coulomb energy of the
system is negligible at initial time.  In Eq.~(\ref{mefi}) the ket
$|\!\! \uparrow \rangle$ indicates spin up state.

To obtain the system evolution we solve the time-dependent Schr\"{o}dinger
equation for the two-particle wavefunction of Eq.~(\ref{mefi}).
Once the real-space
wavefunction is found at a given time step, we compute
the two-particle
density matrix $\rho=|\Psi\rangle\langle\Psi|$ and from the latter  we
calculate the one-particle reduced density matrix $\rho_r$
by tracing over the degrees of freedom of one of the two electrons.
$\rho_r$ is then used to evaluate the entanglement.  In fact, it
is well known that for a
two-fermion system a good correlation measure is given by the von
Neumann entropy of $\rho_r$~\cite{bus, Sch}:
\begin{equation}\label{ricci}
\varepsilon= -\textrm{Tr}[\rho_{r}\ln{\rho_{r}}]=
\sum_{i=1}|z_{i}|^{2}\ln{|z_{i}|^{2}},
\end{equation}
where $|z_{i}|^{2}$ are the eigenvalues of the matrix $\rho_{r}$.

\begin{figure}[htp]
  \begin{centering}
  \includegraphics*[width=0.7\linewidth]{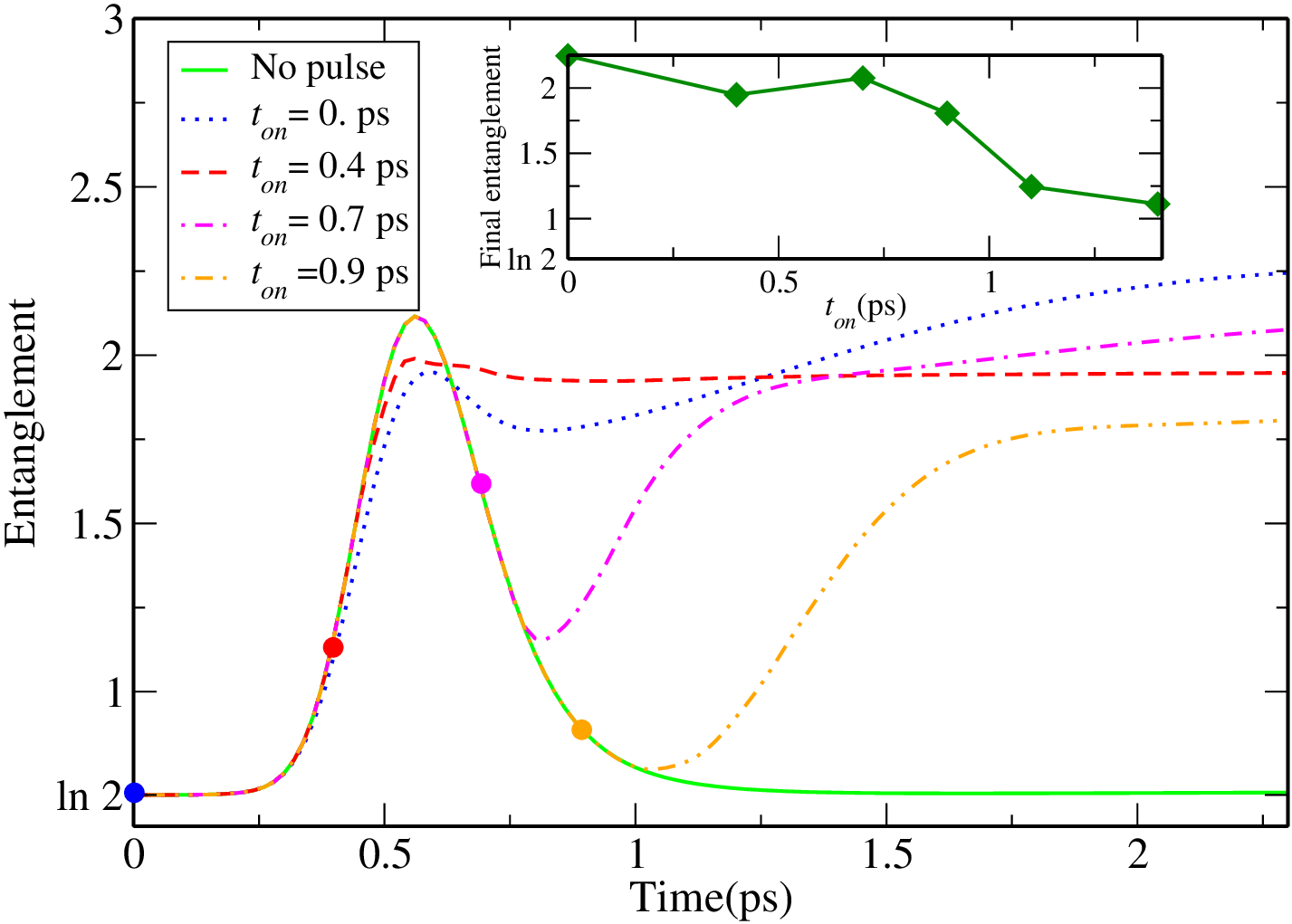}
  \caption{\label{fig2} Entanglement vs. time for
  different   initial times of the pulse $t_{on}$ . Here the pulse duration is  $\Delta\tau_2 = 0.2$ ps. At the initial time the two
  electrons have the same kinetic energy $E_k$=10 meV  corresponding to
  $|k_{in}|=0.290\,\,\textrm{nm}^{-1}$ and are described by two
  wavepackets with mean dispersion $\sigma = 10$ nm moving in opposite
 directions. The filled circle on the curves  indicate the four different $t_{on}$ times. The inset shows
  the stationary values of the entanglement as a function of
  $t_{on}$.}  \end{centering}
\end{figure}

\begin{figure}[htp]
  \begin{centering}
  \includegraphics*[width=0.7\linewidth]{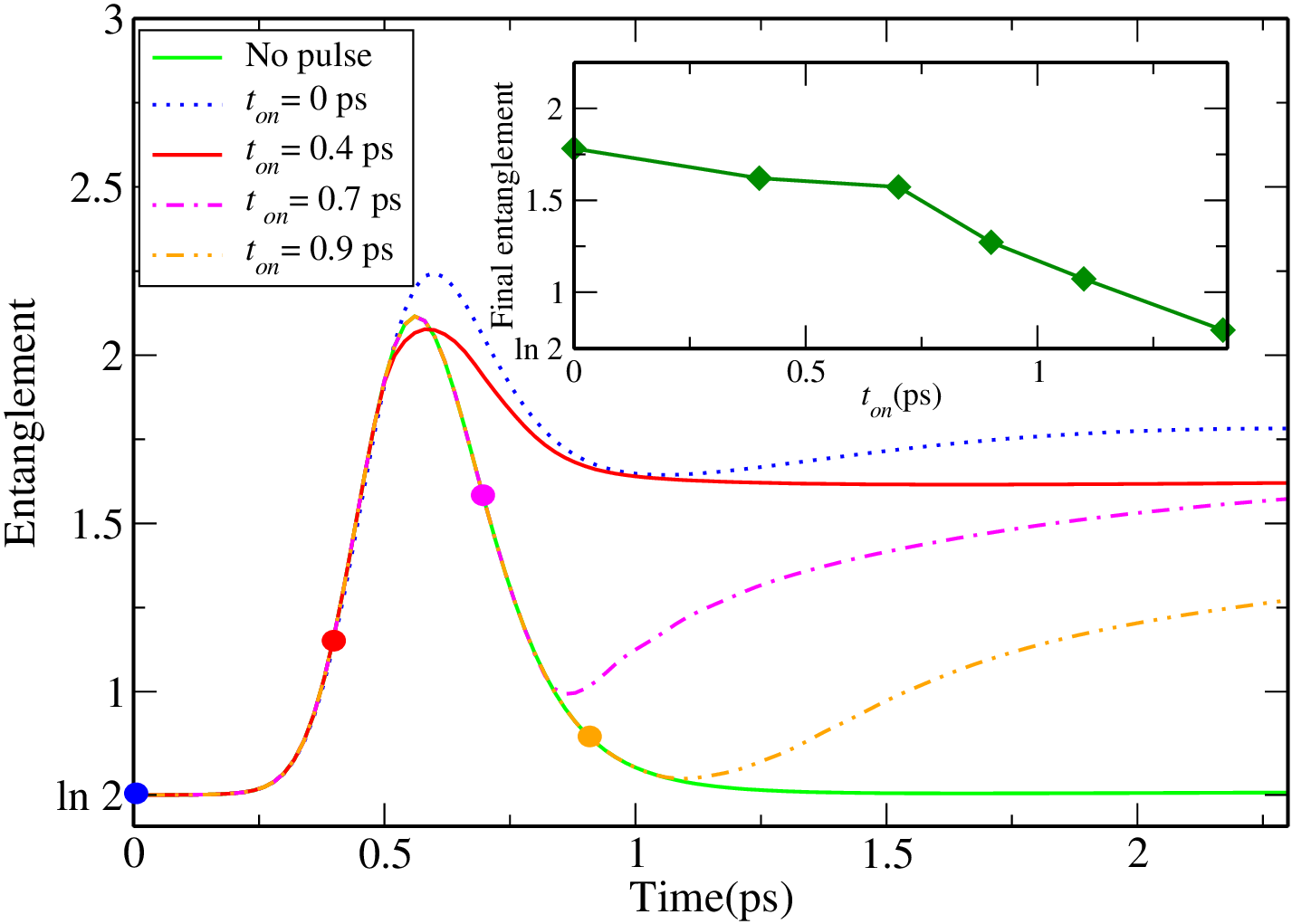}
  \caption{\label{fig3} Same as  Fig.~\ref{fig2}, with a pulse duration  $\Delta\tau_1 = 0.1$ ps.} \end{centering}
\end{figure}

As in the previous section, we consider pulses 0.1~ps and 0.2~ps long.
This implies that each of the two carriers can gain or lose $\hbar
k_{0}$ in its momentum and the corresponding wavefunction splits into
three peaks.  In Fig.~\ref{fig2} and \ref{fig3} we report the time
evolution of the entanglement for $\Delta\tau_1 = 0.1$~ps and $\Delta\tau_2 = 0.2$~ps respectively, for different values of $t_{on}$, i.e.  the time at
which the pulse is switched on.  Since the electron-electron interaction
builds  up quantum correlations in a limited time  interval  (roughly
corresponding to the width of the entanglement peak when no pulse is
present: the solid line in Fig.~\ref{fig2} and \ref{fig3}) it is clear
that by choosing different $t_{on}$ one means to consider  cases with the
potential pulse taking place before ($t_{on}=0$), during ($t_{on}=0.4$
and $t_{on}=0.7$), and after ($t_{on}=0.9$) the scattering.

% \emph{since electron-electron interaction is only effective for a limited time interval (roughly
% corresponding to the width of the entanglement peak when no pulse is
% present: the solid line in Fig.~\ref{fig2} and \ref{fig3}) it is clear
% that by choosing different $t_{on}$ one is considering cases with the
% potential pulse taking place before ($t_{on}=0$), during ($t_{on}=0.4$
% and $t_{on}=0.7$), and after ($t_{on}=0.9$) the scattering.}

At the initial time, the von Neumann entropy is equal to $\ln 2$.
This value is related to unavoidable quantum correlations due to the
exchange symmetry and it does not represent a manifestation of a
genuine entanglement~\cite{bus, Sch}.

\begin{figure}[htb]
  \begin{centering}
  \includegraphics*[width=0.7\linewidth]{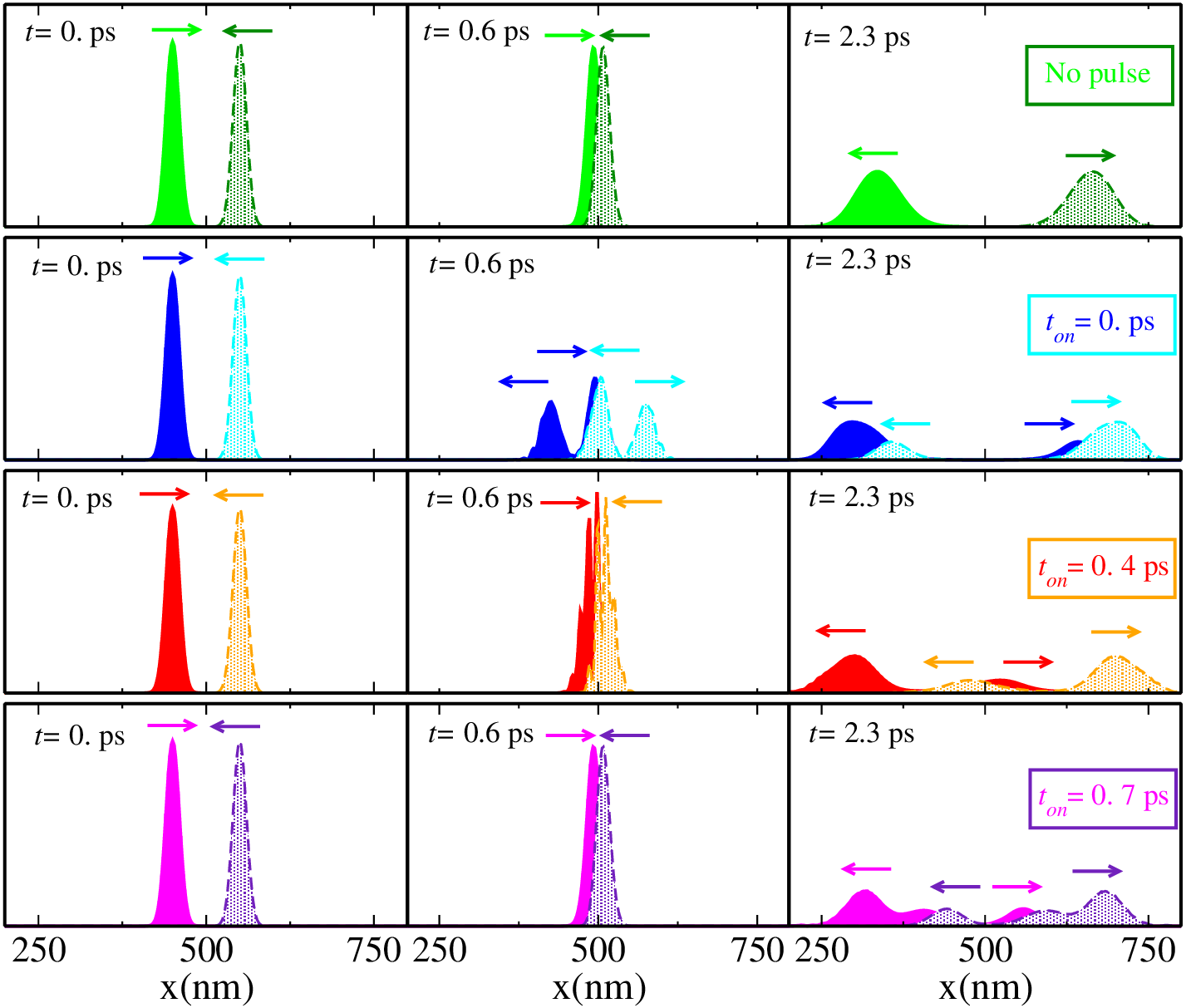}
  \caption{\label{fig4} Particle density at three different time
  steps, namely 0, 0.6 and 2.3 ps, for the case of no pulse( first row) and for
  different values of the initial time of the pulse $t_{on}$, as indicated on the right. Here $\Delta\tau_2 = 0.2$ ps. The solid shaded curve represents the probability density
  $\int |\langle x_a \, x_b|\Psi \rangle|^2 dx_a $ of the
  electron incoming from the left  while the dashed curve describes the
  probability density of the electron incoming from the  right. } \end{centering}
\end{figure}

In absence of pulse, the entanglement increases while the two
electrons are approaching each other, it has a maximum when the
centers of the two wave packets are at the minimum distance, and
finally drops again to the initial value once the electrons get far
apart.  In this case, due to the small thickness of the quantum wire,
the effective Coulomb interaction between the electrons is sufficient
to cause a complete reflection of the two particles  and the
corresponding Gaussian wave packets are reconstructed with opposite
momenta, as can be seen from the top  panel of Fig.~\ref{fig4}.

When the two particles are subject to a pulsed sinusoidal potential,
the entanglement dynamics depends strictly upon the details of the
potential switching on and off.  For example, for $\Delta\tau_2=0.2$ ps and $t_{on}=0$~ps, the peak of the entanglement is lower than
the one found with no pulse.  This is a consequence of the interaction
with the sine-like potential that splits the wave function of each
particle  before the effect of the Coulomb potential becomes
significant.  Thus, the two reflected components will not take part
into the scattering process and will not contribute to the
entanglement formation (see Fig.~\ref{fig4}).  After the peak, for
$t_{on}= 0.4$~ps the entanglement does not vary significantly with
time, while in the case $t_{on}=0$ it appears to increase.  Such a
difference can be ascribed to the fact that the splitting of the
wavefunction can be or be not completed when the Coulomb interaction
gets its maximum value.

For the last two cases of Fig.~\ref{fig2}, $t_{on}=0.7$~ps and
$t_{on}=0.9$~ps, the time evolution of the entanglement turns out to
be quite peculiar.  After the peak (due to the Coulomb scattering) the
entanglement exhibits the same decrease as in the case without 
pulse, up to $t_{on}$, when the pulse is switched on.  At the latter
time, the wave function describing each electron is almost entirely
reflected and splits in various peaks, as described above.  Two of
these, namely the components ``reflected'' by the pulse, propagate in
opposite directions and approach each other.  As a consequence,
Coulomb interaction becomes effective again and gives rise to a second
increase of the entanglement.

We report in Fig.~\ref{fig3} the results for $\Delta\tau_1=0.1$ ps
showing a behaviour similar to the previous case, with few, through
significative, differences.  Specifically, for $t_{on}=0$~ps we
observe that the entanglement shows a peak higher than the one found
in absence of the pulse.  This behavior is different from the one
found for $\Delta\tau_2$ and can be ascribed to the diverse way of
splitting of the two particle wavefunction in the two cases,
 as described in Sec.~\ref{intro}.
 Nevertheless we note that for $t_{on}=0.7$~ps, after the peak (as high
as the one found in absence of the external pulse) the entanglement
decreases until $t_{on}$ and then slowly increases, in qualitative
agreement with  the results  obtained for $\Delta\tau_2$.

Some time after the Coulomb interaction  and the pulse, the entanglement
reaches a stationary value.  This value is displayed in the insets of
the Fig.~\ref{fig2} and \ref{fig3} as a function of $t_{on}$.  We
observe that the largest final entanglement is always found when
$t_{on}=0$.  In this case the wave functions split before the
scattering and both the ``accelerated'' components of the electron and
the central peaks, interact strongly during the Coulomb scattering.

Taking into account the results obtained in absence of pulse, we can
therefore assume that the stationary values of the entanglement depend
upon the transmission in the scattering event: they are greater for
greater transmission probability of the particles.  This is in
qualitative agreement with the results of the theoretical
investigations on the entanglement dynamics in scattering events in 1D
semiconductor structures~\cite{Gun, Bordone}.

\section{Summary and conclusions}\label{intro5}
The growing interest in the entanglement phenomena
\cite{Hars,Law,Tal,bus2,berbor,Oli,Ram} led us to analyze the
electron-electron entanglement dynamics of a two-particle scattering
 in a study  model that
has important applicative perspectives but  whose simplicity makes it of
general interest.  In particular, the appearance of quantum
correlations between  two colliding electrons  is a direct consequence of their
mutual Coulomb repulsion during the scattering event, while the external
periodic potential represents a mean to tailor the electron-electron
interaction and to localize the particles.  We stress  that the
equations of motion of the system under study  are the same that
describe the scattering of electrons by a standing laser
wave\cite{Bate,Min,Efre,Efre2}, where the particles can change their
momentum by an integer multiple of $\hbar k_0$, with $k_0$ the
wave vector of the real-space modulation of the potential.  Our
single-particle time-dependent simulations show directly how the
momentum gain or loss is affected by the switching-on time of the
sinusoidal potential.  In particular the use of pulses in the range of
few tenths of picoseconds leads to a variation of a single $\hbar
k_0$ quantum in the momentum of the particle.

The issues described above allowed us to explain the entanglement
behavior in a two-electron scattering and our two-particle
simulations gave a direct insight on the origin of the correlations.
In fact, we showed that the non-separability of the two-particle state
is mainly originated by the splitting of the spatial wave function
brought about by the interaction with the potential pulse.
 As a matter of fact, our results show that the two particles get more
correlated for longer pulses while entanglement 
dynamics depends closely upon the time of turning on
of the potential.  Moreover, the time of the entanglement
formation, i.e. the time at which the entanglement reaches its
stationary value, is strictly related to the time of the initial
switching-on of the sinusoidal potential.  On the other hand, the
final value of the entanglement depends upon the ratio between the
transmitted and the reflected components of the
wave function\cite{Gun,Bordone}.  We stress that the above
finding shows that the final entanglement will be maximum for the
specific system parameters  that maximize
the splitting of the wave function after the scattering.

Finally,  we note  that the simplicity of the model here investigated
makes it of general interest. In this spirit we think that our
results can be valid guidelines to analyze some phenomena related to
the appearance of quantum correlations in various physical systems
and have important applicative perspectives. The use
 of a sinusoidal potential  could provide a way  to  produce and  control the  electron decoherence due to carrier-carrier collision in low-dimensional semiconductor structures. Moreover our model could be viewed as a prototype system to study the building up of quantum correlations into electron transport assisted by surface acoustic waves through quantum wires since they are modelled as a sinusoidal travelling electric potential which traps the carriers in its moving minima. The results  shown here  could also be helpful to investigate the entanglement formation in the collision event of ions trapped in quasi 1D-structures and subject  to a laser pulse.

\ack The authors would like to  thank Carlo Jacoboni  for useful discussions.
We  acknowledge support from CNR
Progetto Supercalcolo 2008 CINECA. One of the authors (AB)  acknowledges INFM seed project 2008.

\section
*{References}

\end{document}